\documentclass[reprint,
 amsmath,amssymb,
prl,
]{revtex4-1}

\usepackage{graphicx}
\usepackage{filecontents}
\usepackage{tabularx}
\usepackage{dcolumn}
\usepackage{bm}

\begin{document}

\preprint{APS/123-QED}

\title{Charge Density Modulation and Defect Ordering in Na$_x$MnBi$_y$ magnetic semimetal}

\author{A. Wegner}
\affiliation{Department of Physics, University of Virginia, Charlottesville, VA 22904, USA}
\author{D. Louca*}
\affiliation{Department of Physics, University of Virginia, Charlottesville, VA 22904, USA}
\author{K.M. Taddei}
\affiliation{Neutron Scattering Division, Oak Ridge National Laboratory, Oak Ridge, TN 37831, USA}
\author{J. Neuefeind}
\affiliation{Neutron Scattering Division, Oak Ridge National Laboratory, Oak Ridge, TN 37831, USA}

\date{\today}

\begin{abstract}
The I-Mn-V antiferromagnet, NaMnBi, develops a very large positive magnetoresistance (MR) up to 10,000 \%\ at 2 K and 9 T (see Ref.\cite{yang2018defect}) in crystals showing a semiconductor-to-metal transition (SMT). In the absence of an SMT, a modest (20 \%) MR is achieved. Here, we show that upon cooling below the magnetic transition, a spatial modulation appears giving rise to new Bragg peaks due to charge and defect ordering in a checkerboard pattern, with two kinds of modulation vectors, $q_1$=($\frac23$, 0, 1) and $q_2$=($\frac23, \frac13, \frac12$). This constitutes a superlattice transition ($T_s$) that lowers the symmetry from the high temperature centrosymmetric P4/nmm to the non-centrosymmetric P$\overline4$m2. In crystals with a large MR, a close to room temperature $T_s$ is observed with $q_1$ appearing first, followed by $q_2$. In crystals with low MR however, $T_s$ is much lower  and only $q_1$ is observed. The charge modulation and spin fluctuations may both contribute to the enhancement of MR.

\end{abstract}

\maketitle

Semiconducting electronics have dramatically evolved over the decades in part due to advances in fundamental research and engineering that led to quick turnarounds from discovery to commercialization\cite{prinz}. A well-known example is spin polarized transport which applies a phenomenon referred to as giant magnetoresistance (GMR)\cite{wolf2001spintronics, reig2013giant, willoughby2015spintronics} in read heads and magnetic information storage\cite{engel20054}. Commercial spintronic, electronic and optoelectronic devices are largely based on ferromagnetically ordered multilayer films exhibiting GMR \cite{chappert, loth, moser}. What is the mechanism that yields the large MR? Classical MR is a weak effect that commonly appears in non-magnetic systems under a magnetic field, H. It is usually a negative quantity, associated with the orbital motion of the charge carriers and follows a quadratic field dependence at low fields that saturates quickly with an increasing field\cite{ishiwata}. On the other hand, the appearance of an extremely large electrical response under field is unusual and positive, extreme (XMR) has been observed that is orders of magnitude stronger than GMR. XMR exhibits a more complex magnetic field dependence that may be largely driven by the intrinsic electronic structure. First observed in the nonmagnetic and nearly compensated semimetal WTe$_2$, \cite{ali} the mechanism behind XMR is still an open question with spin-lattice-charge and orbital correlations playing an important role. Since the initial observation in transition metal dichalcogenides, XMR has been observed in a wide class of binaries \cite{liang, tafti, wang, wu}.

New research directions in the search for promising candidates for the next generation of MR devices is on the rise. This includes magnetic and non-magnetic quasi-two dimensional (2D) and three-dimensional (3D) semiconductors \cite{awschalom, lin}. Currently, there is immense interest in expanding applications of antiferromagnets (AFMs) in spintronics \cite{bhattacharyya2018high}. A particularly interesting group of AFM materials is the I-Mn-V class of semiconductors with tetragonal crystal lattices. Materials in this class are layered and can easily be exfoliated. LiMnAs has been shown to be compatible with semiconducting substrates\cite{jungwirth2011demonstration}. The isostructural CuMnAs, when stabilized in a tetragonal phase by a lattice matched substrate has been used in prototype memory devices\cite{vzelezny2018spin}. CuMnAs devices are made possible because of a locally broken inversion symmetry that allows magnetic writing by current pulses with the inverse spin galvanic effect while the device can be read by anisotropic MR \cite{olejnik2017antiferromagnetic}.

In this work, we focus on NaMnBi, a semiconductor with $T_N$ = 340 K. Its crystal structure is shown in the inset of Fig. 1(b), with C-type magnetic ordering. The Mn spins at the corners are aligned ferromagnetically and only the center spin is anti-aligned, creating an overall AFM order. The crystal symmetry is nominally tetragonal with the $P4/nmm$ space group\cite{bronger1986magnetic}. The Mn atoms are tetrahedrally coordinated by Bi and a Na layer separates the Mn-Bi layers as seen from the crystal structure. It was recently demonstrated that a giant MR emerges in quenched (Q) samples of NaMnBi that carry more defects than the slowly furnace-cooled (FC) samples \cite{yang2018defect}, indicating that quenched disorder changes the fundamental properties of the system. The FC and Q-crystals carry a different concentration of defects. The FC crystals are closer to a stoichiometric composition and show no SMT in zero-field. The MR reaches 20\% at 2 K and 9 T and 7\% at room temperature. The quenched crystals host more vacancies and show an SMT with metallic transport at low temperatures \cite{yang2018defect}. The MR rises to 10,000 \% at 2 K and 9 T and 600\% at room temperature, rendering  NaMnBi the first material in the I-Mn-V class to exhibit a notably high MR.
In this Letter, we report that coincidental with the SMT, a structural transition ($T_S$) to a new symmetry, $P\overline{4}m2$, occurs upon cooling, due to the ordering of defects and charges. The transition to the new symmetry is accompanied by a 3 x 3 x 1 modulation of the unit cell with a displacement vector $q_1$=($\frac23$, 0, 1). $T_S$ occurs below 300 K in Q-crystals and below 100 K in FC-crystals. A second modulation vector is observed in Q-crystals only upon further cooling with a 3 x 3 x 2 cell expansion and a $q_2$=($\frac23, \frac13, \frac12$). 


%

Single crystals of NaMnBi were grown using the Bi self-flux method. Details of the sample growth are provided in Ref. \cite{yang2018defect}. For the neutron experiments at HB-2A and the Nanoscale Ordered Materials Diffractometer (NOMAD) at Oak Ridge National Laboratory (ORNL) the crystals were crushed just prior to the experiment. Quenched and furnace cooled crystals were measured at T = 4, 10, and between 50-350 K in steps of 50 K. The measurements at HB-2A were performed using a $\lambda$ = 2.411 \AA\. The quenched crystal composition is estimated to be Na$_{0.82}$MnBi$_{0.86}$ from energy dispersive X-ray spectroscopy while the furnace cooled sample has a composition of Na$_{0.92}$MnBi\cite{yang2018defect}. Quenching leads to a high MR. 


In Figs. 1(a) and 1(b), diffraction data collected for the Q- and FC-crystals are shown as a function of temperature. At 350 K, the data are fit using the published P4/nmm crystal symmetry\cite{bronger1986magnetic}. Both samples contain a second phase of pure Bi. At 350 K, although no magnetic peaks can be discerned in the diffraction pattern for the FC-crystal, very weak magnetic peaks are seen in the diffraction pattern of the Q-crystal. By 300 K, the magnetic peaks become clearly visible in both samples. In Fig. 1 (c) the order parameter of the magnetic transition is shown. The order parameter is determined from the integrated intensity of the (1, 0, 0) magnetic Bragg peak. The intensity is normalized to the integrated intensity of the (1, 1, 0) nuclear peak, as it is the strongest peak with no temperature dependence. The magnetic transition is sharper in the Q-crystal, in spite of the increased number of defects. The magnitude of the magnetic moment is approximately 4.5 $\mu_B$, and is the same for both crystal types. 

Upon further cooling, a structural transition is observed for the first time. In the Q-crystal, peaks appear at 2$\theta_1$=27.2$^\circ$  and 2$\theta_2$ = 24.5$^\circ$ as indicated in Fig. 1(a) (d-spacing = 5.07 \AA\ and 5.65 \AA\, respectively), while in the FC-crystal only the 2$\theta_1$ peak is observed (Fig. 1(b)). The onset temperature, $T_s$, is below 300 K in the Q sample and below 100 K in the FC sample. Fig. 1(d) is a plot of the order parameter for $T_s$ based on the temperature dependence of displacement vector $q_1$. These peaks cannot be indexed using the high temperature symmetry of P4/nmm. The strongest peak is the 2$\theta_1$ which is clearly visible in both samples, and can be indexed to a 3 x 3 x 1 supercell, with an index of $q_1$=($\frac23$, 0, 1). The weaker, 2$\theta_2$ observed only in the Q-crystal is indexed to a $q_2$=($\frac23, \frac13, \frac12$) with a 3 x 3 x 2 lattice modulation. $q_2$ is discernible only in the 4 K data (see Supplemental Fig. 1S). Thus two superstructures with two characteristics q-vectors are present in Q-crystals that might either arise from one domain or from multiple domains. The strongest superlattice has no doubling in the c-direction, while the weaker superlattice has doubling along c, expanding the cell in all three directions. For the Q-crystal, the ($\frac23$, 0, 1) peak has an integrated intensity that is 10 times stronger than in the FC sample. The temperature at which the superlattice sets in coincides with a change in the bulk susceptibility ($\chi$) as well as a step down in the transport (see Ref.\cite{yang2018defect}) which is reminiscent of the effects observed in other materials showing charge density wave (CDW) transitions such as in TiSe$_2$ \cite{di1976electronic, wegner2019local, wegner2018evidence}.

\begin{figure}[t!]
\includegraphics[width=3.4 in]{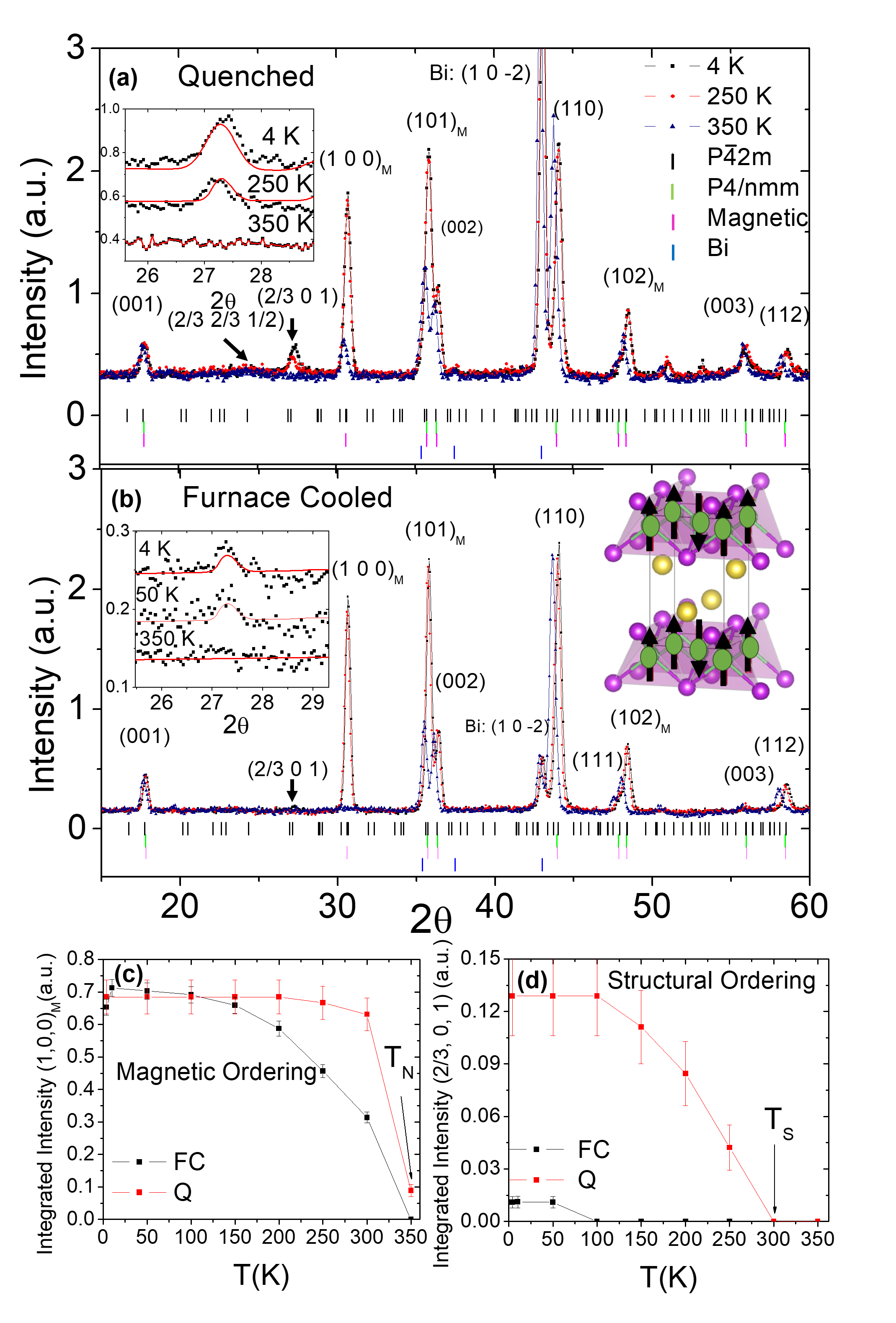}
\caption{\label{diagram}(a) The temperature dependence of the neutron diffraction data collected at HB2A of Q-NaMnBi show a structural phase transition with a T$_s$ around 250 K. The $q_1$=(2/3, 0, 1) peak fit by the model is shown in the inset. (b) The structural transition in the FC-NaMnBi has a critical temperature below 100 K and is much weaker than in the quenched sample. The order parameter for the magnetic (c) and structural (d) transitions are shown. The order parameter is the normalized integrated intensity of the (2/3, 0, 1) structural and the (1, 0, 0) magnetic peaks.}
\end{figure}

The symmetry that can reproduce the superlattice reflections is $P\bar4m2$. This symmetry breaks inversion of the high temperature phase. In the FC crystal, the 3 x 3 x 1 superlattice consists of modulations of Bi ions as shown in the $ab-$plane projection of Supplemental Fig. 3S. For the FC-crystal, all Mn ions have tetrahedral coordination. The projection of the structure in the $ab-$plane with the $P\bar4m2$ for the Q-crystal is shown in Fig. 2. Note that this model is obtained from the Rietveld refinement of the diffraction data. The $P\bar4m2$ calls for ordering the vacancies in a checkerboard stripe pattern as shown in the figure. Four different Bi coordination environments are present, indicating that the Bi-sublattice is distorted. The Bi-Mn coordinated polyhedra shown in Fig. 2(b-e) are color coded with the same color scheme shown in the projected structure on the left. The Bi atoms are displaced by as much as 0.05 \AA\ from their undistorted high temperature position. The Na atoms tend to distort in the opposite direction to that of Bi, while the Mn ions remain unchanged. In the Q-crystal, the Bi site located at $(0, \frac12, z)$ and $(\frac12, 0, z)$ has a refined occupancy of only 0.57 (see Table II in the supplement). When a Bi is missing from the tetrahedron, the Bi adjacent to the vacancy along the a- or b-axis switches to the opposite layer. This leads to the large change in the intensity of the ($\frac23, 0, 1$) peak. It also changes the coordination geometry of some of the Mn-Bi polyhedra as shown in Fig. 2. The blank parts with no polyhedra coloring correspond to the locations of the Bi vacancies. Due to the vacancies and layer switching of the Bi, there are in addition to the tetrahedral unit shown in (b), other geometries such as the rectangular planar shown in (c) and two different trigonal pyramidal local environments as well shown in (d) and (e). As $T_N$ is above $T_s$, it is inferred that the distortion of the magnetic Mn sublattice is negligible, and the Mn ions occupy special positions in the unit cell. However, as shown in Fig. 2(f), the Mn thermal factor obtained from the Rietveld refinement is significantly higher than that for Bi or Na and much higher in the Q-crystal than in the FC-crystals which suggests that there is a large uncertainty in the position of the Mn ions that picks up below the magnetic transition.

\begin{figure}[t!]
\includegraphics[width=3.4 in]{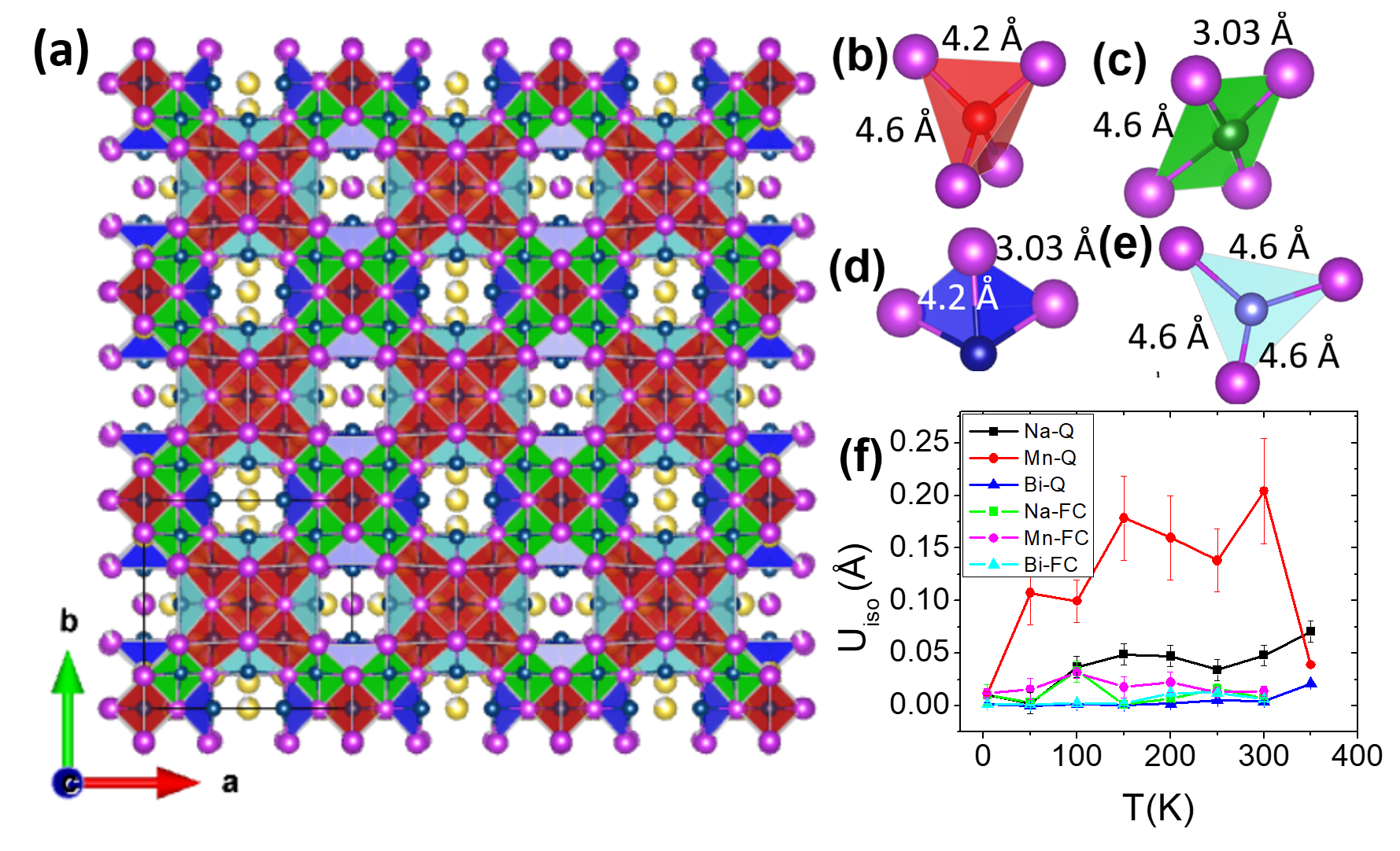}
\caption{\label{diagram}(a) The 3x3 unit cell obtained from the Rietveld refinement of the low temperature data for Q-NaMnBi. Na is in yellow. Mn is at the center of the polyhedron. Bi is in purple. The polyhedra are color coded. The polyhedra in (b,c,d,e) show the different coordination of Bi displayed (a).(f) The thermal parameters from Rietveld refinement are plotted as a function of temperature for the Q and FC samples.}
\end{figure}

Comparison of the real-space arrangement of the FC- and Q-defect structures is enabled by Fourier transforming the total structure function from the neutron data of Fig. 1 (see Ref. \cite{egami2003underneath}). A $Q_{max}$ = 37$^{-1}$ \AA\ was used for the transform to obtain the pair correlation function, G(r). In Fig. 3, the G(r) is plotted as a function of distance (r) between pairs of atoms. The peak intensity corresponds to the probability of finding a particular pair in space. Negative peaks involve Mn correlations with other atoms, and they are negative because Mn has a negative neutron scattering length \cite{sears1986neutron}. In Fig. 3(a), the Q-NaMnBi experimentally obtained G(r) (shown in symbols) at 5 K is compared to two model G(r) functions: one corresponding to the average model of the high temperature P4/nmm phase alone (plotted with a green line) and the other corresponding to a model constructed by combining 57\% of the P4/nmm phase with 43 \% of Bi metal, the latter of which is the second phase determined from the Rietveld refinement. The Q-crystals contain significantly more excess Bi than the FC-crystals. The G(r) of pure Bi metal is shown using a blue line in the same plot. From the G(r) of pure Bi, the Bi-Bi correlations from the second phase can be identified. It can be seen that while some features in the experimental G(r) can be reproduced by the model G(r) with only the P4/nmm phase, adding Bi as is done in the total G(r) model (red line) improves the fit somewhat but the overall agreement is still not good ($\chi^2$ = 0.901). In the high temperature phase, all Mn ions are tetrahedrally coordinated by Bi. The first Mn-Bi peak is at 2.8 \AA, and is reproduced by the model. Beyond the first peak though the model does not fit the data well. Similarly, for the FC-NaMnBi data shown in Fig. 3(b), the fitting with the high temperature P4/nmm symmetry combined with excess Bi does not reproduce the data well either ($\chi^2$ = 0.184). Less excess Bi is present (18 \%) in the total G(r) in this case, and the peaks are not as intense, but the overall fitting fails to reproduce the local structure in the 3.0-3.5 \AA\ range. 

On the other hand, upon using the P$\overline4$m2 symmetry to calculate a model G(r), the comparison to the same experimental data is far better for both samples. The results are shown in Figs. 3(c-d). Shown in Fig. 3(c) are the experimental G(r) obtained from the data for Q-NaMnBi compared to a model G(r) calculated from the refined parameters of the $P\bar4m2$ phase alone (green line) and to a model that takes excess Bi (red line) into account. The agreement is quite good with a $\chi^2$ = 0.086. In this model, due to the Bi switching layers, Bi-Bi bonds at 3.03 \AA\ are present that can now reproduce the data well in that range. Moreover, the Na-Bi bonds can now fit the peak at 3.49\AA\ well. Similarly in Fig. 3(d), the FC-NaMnBi data is compared with the G(r) calculated from the $P\bar4m2$ model alone (green line) and with the addition of excess Bi (red line). It, too, shows as good of an agreement as in the previous sample. The agreement factor in this case is $\chi^2$ = 0.101.
How different is the local environment between the Q- and FC-crystals? This is shown in Fig. 3(e) which is a comparison of the total model G(r) of Fig. 3(c) and 3(d) but with the excess Bi subtracted. Shown with a red line is the G(r) model of the FC-crystal used in the fitting of Fig. 3(d) and, similarly, the blue line is the G(r) model of the Q-crystal used in the fitting in Fig. 3(c). The two models are quite similar overall but with differences between the Na-Bi and Bi-Bi correlations in the 3.0-3.5 \AA\ range. Short Bi correlations are observed even in the FC-crystal, although there are fewer short Bi-Bi bonds than in the Q-crystal by comparing the intensity. This is due to a smaller number of vacancies, so that although on average all Mn ions have tetrahedral coordination in the FC-sample, the few vacancies that do exist affect their neighbors in the same way as the vacancies in the Q-sample. Also shown in this figure is a calculated G(r) model for the high temperature phase, P4/nmm. This clearly shows that the local structure below T$_s$ is substantially different from the high temperature phase with noticeable differences i.e. the first Bi-Bi correlation should appear at 4.5 \AA\ in the high temperature symmetry, but instead it is at 3.03 \AA\ (this distances is due to the  rectangular planar and trigonal prismatic polyhedra shown in Figs. 2(c) and 2(d)), and similarly for the Na-Bi correlations at 3.49 \AA.

\begin{figure}[t!]
\includegraphics[width=3.4 in]{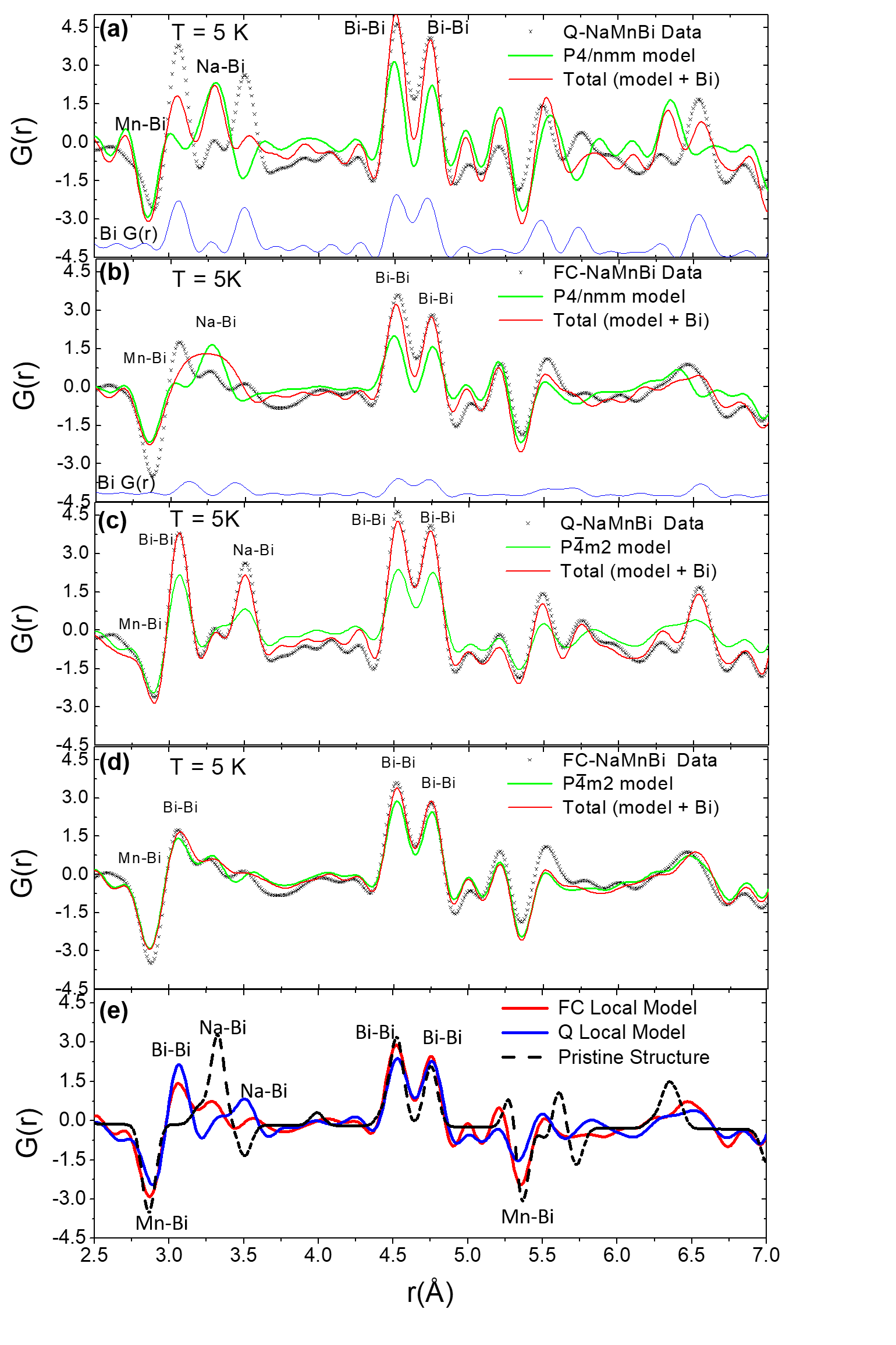}
\caption{\label{diagram} The PDF for the (a) Q crystals and (b) FC crystals are fit with the P4/nmm symmetry at 5 K. Some features in the local structure are not reproduced even when 18\% and 43\% of Bi are included in the Q and FC crystals, respectively. The PDF for the (c) Q crystals and (d) FC crystals are fit with the $P\overline{4}m2$ symmetry at 5 K. (e) The local models of the FC and Q crystals are compared to the model for the pristine P4/nmm symmetry.}
\end{figure}

Despite differences in stoichiometry, the Q- and FC-crystals have a magnetic order with a large magnetic moment of 4.5 $\mu_b$ and a similar $T_N$. The presence of the periodic lattice distortions and vacancy ordering leads to a superlattice modulation with broken inversion symmetry and a charge density modulation. This may be crucial to controlling spin-polarized electrons with spin-orbit torque, an important property in AFM spintronic applications\cite{jungwirth2016antiferromagnetic}. While we cannot decide on a particular mechanism that describes the charge modulations, the results suggest that electron-phonon coupling is important. Two factors are contributing to the MR: 1) charge modulation and 2) spin scattering. A weak MR is observed in FC crystals and is related to a weak T{$_s$} transition, while a very large MR is observed in Q-crystals with a high T{$_s$} transition temperature and defect ordering. Moreover, even though T{$_N$} is robust against doping, the Mn thermal factor shown in Fig. 2(f) is substantially larger in Q-crystals which most likely contributes to spin scattering. Although it remains difficult to confirm the exact mechanism driving the MR, the material is promising for further study.

The authors would like to thank C. dela Cruz and M. Tucker for their assistance with the neutron experiments. They would also like to acknowledge support from the Department of Energy, Grant No. DE-FG02-01ER45927. A portion of this research used resources at the High Flux Isotope Reactor and the Spallation Neutron Source, a DOE Office of Science User Facility operated by the Oak Ridge National Laboratory. 

*To whom correspondence should be addressed: louca@virginia.edu


\end{document}